\documentclass[a4paper]{article}
\usepackage{amsfonts}

\usepackage{graphicx}
\usepackage{amsmath}

\begin{document}

\title{Distortion operator and Entanglement Information Rate Distortion of Quantum
Gaussian Source}
\author{Xiao-yu Chen \\
Lab. of Quantum Information, China Institute of Metrology, Hangzhou, 310018,
China;}
\date{}
\maketitle

\begin{abstract}
Quantum random variable, distortion operator are introduced based on
canonical operators.  As the lower bound of rate distortion, the
entanglement information rate distortion is achieved by Gaussian map for
Gaussian source. General Gaussian maps are further reduced to unitary
transformations and additive noises from the physical meaning of distortion.
The entanglement information rate distortion function then are calculated
for one mode Gaussian source. The rate distortion is accessible at zero
distortion point. For pure state, the rate distortion function is always
zero. In contrast to the distortion defined via fidelity, our definition of
the distortion makes it possible to calculate the entanglement information
rate distortion function for Gaussian source.
\end{abstract}

\textit{Distortion operator: }Two major parts in classical information
theory are channel capacity and rate distortion theory. They concern
respectively with the reliability and effectiveness of information
transmission. In quantum information theory, channel capacity has been
widely investigated, but little effort has been put into developing quantum
rate-distortion theory\cite{Barnum}\cite{Devetak}.\textit{\ }It was proven
\cite{Barnum} that the quantum rate-distortion function $R(D)$ is lower
bounded by entanglement information rate-distortion function $R^I(D).$ For a
given source $R^I(D)$ is defined by
\begin{equation}
R^I(D)=\min_{\mathcal{E}\left| d(\mathcal{E)\leq D}\right. }I_c(\rho ,%
\mathcal{E}).
\end{equation}
where $d$ is some distortion function, and $\mathcal{E}$ is the channel. The
result is proven under the assumption of distortion function defined by
transmission fidelity. It is linear among different modes. We can extend the
distortion function to a more general form. The result will also be true if
it is linear among different modes. One of the useful distortion function is
mean square function as used in classical information theory for Gaussian
source. The mean square distortion in classical theory is $\overline{d}%
=E(d(Y,Y^{\prime }))=\int p(y,y^{\prime })d(y,y^{\prime })dydy^{\prime }$,
with $d(y,y^{\prime })=(y-y^{\prime })^2.$ Where $Y$ is the input random
variable and $Y^{\prime }$ is the output, $p(y,y^{\prime })$ is the joint
density distribution function. The same idea should be extended to quantum
information theory. What is the quantum corresponding of random variable? We
prefer the canonical operators $X$ and $P$. Then the distortion operator
will be introduced as
\begin{equation}
d(A,B)=\frac 12\sum_{i=1}^n[(X_{Ai}-X_{Bi})^2+(P_{Ai}+P_{Bi})^2].
\end{equation}
Where $A$ is the sender and $B$ is the receiver. The Schmidt purification of
the sender state $\rho _A$ is obtained by introducing the reference system,
described by Hilbert space $\mathcal{H}_R$, isomorphic to the Hilbert space $%
\mathcal{H}_Q=\mathcal{H}_A$ of the initial system, Then there exists a
purification of the state $\rho _A,$ a unit vector $\left| \psi
\right\rangle \in \mathcal{H}_Q\otimes \mathcal{H}_R$ such that $\rho
_A=Tr_R\left| \psi \right\rangle \left\langle \psi \right| .$Where$\left|
\psi \right\rangle =\sum_k\sqrt{\lambda _k}\left| \lambda _k\right\rangle
\left| \lambda _k\right\rangle $, with $\lambda _k$ and $\left| \lambda
_k\right\rangle $ are eigenvalue and eigenvector of $\rho _A$ respectively.
After transmission, the joint state will be $\rho ^{RQ^{\prime }}=(\mathcal{E%
}\otimes \mathbf{I})\left| \psi \right\rangle \left\langle \psi \right| ,$
it is easy to verify that $\rho _A=Tr_Q(\rho ^{RQ^{\prime }})$ and $\rho
_B=Tr_R(\rho ^{RQ^{\prime }})=\mathcal{E}(\rho _A),$ which means the system $%
R$ remains at the input state $\rho _A$ and system $Q$ evolves to the output
state $\rho _B.$ The average distortion will be
\begin{equation}
\overline{d}=Tr\rho ^{RQ^{\prime }}d(A,B).
\end{equation}
A similar quantity was introduced to obtain entanglement of formation of
symmetric Gaussian states \cite{Giedke1}. The average distortion possesses
some kind of EPR-uncertainty of the joint state, we here neglect the mean of
canonical operators for simplicity, thus all of the first moments of the
states will be neglected in the follows.

\textit{Entanglement information rate-distortion function: }The coherent
information $I_c(\rho ,\mathcal{E})=S(\mathcal{E}(\rho ))-S(\rho
^{RQ^{\prime }}).$ The Entanglement information rate-distortion function is
the global minimum of the coherent information under the certain distortion.
There is a useful lemma in classical information theory which gives
necessary and sufficient conditions for the global minimum of a convex
function of probability distributions in terms of the first partial
derivatives. The lemma was extended to quantum information theory \cite
{Holevo} in evaluating the capacities of bosonic Gaussian channels. Let $F$
be a convex function on the set of density operators which contains $\rho _0$
and $\rho ,$ the necessary and sufficient condition for $F$ achieves minimum
on $\rho _0$ is that the convex function $F((1-t)\rho _0+t\rho )$ of the
real variable $t$ achieves minimum at $t=0$ for any $\rho .$ That is $\frac
d{dt}\left| _{t=0}F((1-t)\rho _0+t\rho )\geq 0\right. .$ Here $I_c(\rho ,%
\mathcal{E})$ is a function of density operator $\rho ^{RQ^{\prime }}.$
Coherent information is convex due to channel operation\cite{Barnum}, that
is for operation $\mathcal{E}_\lambda \equiv \lambda \mathcal{E}_1\mathcal{+}%
(1-\lambda )\mathcal{E}_2,$ where $0\leq \lambda \leq 1,$ one has $I_c(\rho ,%
\mathcal{E}_\lambda )\leq \lambda I_c(\rho ,\mathcal{E}_1)+(1-\lambda
)I_c(\rho ,\mathcal{E}_2)$. Thus coherent information is a convex function
of density operator $\rho ^{RQ^{\prime }}.$ Suppose the minimum is achieved
at $\rho _0^{RQ^{\prime }}=(\mathcal{E}_0\otimes I)\left| \psi \right\rangle
\left\langle \psi \right| ,$ the necessary and sufficient condition will be
\begin{equation}
\frac d{dt}\left| _{t=0}I_c(\rho ,(1-t)\mathcal{E}_0+t\mathcal{E})\geq
0\right. .  \label{wave0}
\end{equation}
The derivative will be$-Tr(\mathcal{E}(\rho )\mathcal{-E}_0(\rho ))\log
\mathcal{E}_0(\rho )+Tr(\rho ^{RQ^{\prime }}-\rho _0^{RQ^{\prime }})\log
\rho _0^{RQ^{\prime }}.$ If $\mathcal{E}_0$ is a trace preserving completely
positive (CP) Gaussian operation, then for a Gaussian input state $\rho $,
the output state $\mathcal{E}_0(\rho )$ and the joint state $\rho
_0^{RQ^{\prime }}$ will be Gaussian. Hence their logarithms are quadratic
polynomials in the corresponding canonical variables\cite{Holevo}. The
derivative will be zero under the constrains of the first and second
moments. Where the trace preserving property of $\mathcal{E}$ is also used.
The conclusion is that for any channels with the same first and second
moments, Gaussian channel achieves the minimum of coherent information. The
moments of the channel is with respect to a given Gaussian input state.

\textit{Gaussian channel: }Gaussian CP maps are defined as maps which
transform Gaussian states into Gaussian states. Gaussian CP map is thus
isomorphic to bipartite Gaussian state\cite{Giedke2}\cite{Fiurasek}
\begin{equation}
G=\int_{\Bbb{R}^{4n}}dx\exp (-\frac 14x^T\Gamma x+iD^Tx-C)W(x),
\end{equation}
where $W(x)=\exp [-ix^TR]$ are Weyl operators and $R=(X_1,P_1,X_2,\cdots
,P_{2n}),$ with $[X_k,P_l]=i\delta _{kl}$.We in the following will omitted
the linear part $D^T$ and the constant $C$ which are not critical in our
problem. The output state will be $\mathcal{E}(\rho )\propto
Tr_2[G^{T_2}\rho ],$where the trace is taking on the second part of $G^{T_2}$
and the input state $\rho $. The completely positive map on the input state
will be $\rho ^{RQ^{\prime }}=(\mathcal{E}\otimes \mathbf{I)(}\left| \psi
\right\rangle \left\langle \psi \right| \mathbf{).}$ The correlation matrix
(CM) of the Schmidt purification $\left| \psi \right\rangle $ is \cite
{Holevo}
\[
\gamma _\psi =\left[
\begin{array}{ll}
\gamma & \beta \\
\beta ^T & \gamma
\end{array}
\right] ,
\]
where $\gamma $ is the CM of input state $\rho $, $\beta =-\beta ^T=J_n\sqrt{%
-(J_n^{-1}\gamma )^2-\mathbf{I}}$ are purely off-diagonal$,$ where
\[
J_n=\bigoplus_{k=1}^nJ,\text{ }J=\left[
\begin{array}{ll}
0 & -1 \\
1 & 0
\end{array}
\right]
\]
Every operators $A\in \mathcal{B}\mathcal{(H)}$ is completely determined by
its characteristic function $\chi _A(x):=Tr[AW(x)]$ \cite{Petz}. It follows
that $A$ may be written in terms of $\chi _A$ as\cite{Perelomov} $A=\pi
^{-m}\int_{\Bbb{R}^m}dx\chi _A(x)W(-x).$ Thus$\left| \psi \right\rangle
\left\langle \psi \right| =\pi ^{-4n}\int_{\Bbb{R}^{4n}}dx\chi _\psi
(x)W(-x),$ with $\chi _\psi (x)=\exp [-\frac 14x^T\gamma _\psi x]$, we
assume the first moments of the input state $\rho $ be zero. Hence $\rho
^{RQ^{\prime }}\mathbf{=}\pi ^{-4n}\int_{\Bbb{R}^{4n}}dx_1dx_2\chi _\psi
(x_1,x_2)W(-x_2)Tr_2[G^{T_2}W(-x_1)]$The CM of $\rho ^{RQ^{\prime }}$ will
be
\begin{equation}
\left[
\begin{array}{ll}
\widetilde{\Gamma }_1-\widetilde{\Gamma }_{12}(\widetilde{\Gamma }_2+\gamma
)^{-1}\widetilde{\Gamma }_{12}^T & \widetilde{\Gamma }_{12}(\widetilde{%
\Gamma }_2+\gamma )^{-1}\beta \\
\beta ^T(\widetilde{\Gamma }_2+\gamma )^{-1}\widetilde{\Gamma }_{12}^T &
\gamma -\beta ^T(\widetilde{\Gamma }_2+\gamma )^{-1}\beta
\end{array}
\right] .  \label{wave1}
\end{equation}
Where we have denoted
\[
\Gamma =\left[
\begin{array}{ll}
\Gamma _1 & \Gamma _{12} \\
\Gamma _{12}^T & \Gamma _2
\end{array}
\right] ,
\]
and $\widetilde{\Gamma }=(\mathbf{I\oplus }\Lambda )\Gamma (\mathbf{I\oplus }%
\Lambda )$, with $\Lambda =diag(1,-1,1,-1,\cdots ,1,-1)$ is a diagonal
matrix which represents the transposition in phase space $(X_j\rightarrow
X_j,P_j\rightarrow -P_j).$The CM of the out output state $\mathcal{E}(\rho )$
is $\gamma ^{\prime }=\widetilde{\Gamma }_1-\widetilde{\Gamma }_{12}(%
\widetilde{\Gamma }_2+\gamma )^{-1}\widetilde{\Gamma }_{12}^T$. Now we turn
to the trace-preserving Gaussian CP maps\cite{Demoen} which describe all
actions that can be performed on $\rho $ by first adding ancillary systems
in Gaussian states, then performing unitary Gaussian transformations on the
whole system, and finally discarding the ancillas. On the level of CMs these
operations were shown to be described by $\gamma \mapsto \gamma ^{\prime
}=M^T\gamma M+N.$ The Gaussian operator that corresponds to this operation
has the CM\cite{Giedke2}
\[
\Gamma =\lim_{r\rightarrow \infty }\left[
\begin{array}{ll}
M^TA_rM+N & M^TC_r \\
C_rM & A_r
\end{array}
\right] ,
\]
where $A_r=\cosh r\mathbf{I}$ and $C_r=\sinh r\Lambda $. By taking the
limitation of $r\rightarrow \infty $ one has the CM of $\rho ^{RQ^{\prime }}$
to be
\begin{equation}
\Gamma ^{\prime }=\left[
\begin{array}{ll}
\Gamma _1^{\prime } & \Gamma _{12}^{\prime } \\
\Gamma _{12}^{\prime T} & \Gamma _2^{\prime }
\end{array}
\right] =\left[
\begin{array}{ll}
M^T\gamma M+N & M^T\beta \\
\beta ^TM & \gamma
\end{array}
\right] .  \label{wave2}
\end{equation}
This is the final result for a trace-preserving Gaussian map on an input
Gaussian state. The reason that we restrict ourself to the trace-preserving
Gaussian CP maps is from the physical consideration. The result state of
trace-preserving Gaussian CP map is a state with its CM $\gamma $ remains
intact in the reference system $R$ (see Eq.(\ref{wave2})), hence we can
compare the output of the $Q$ system with the input state which is keep
intact in $R$ system. While a general Gaussian CP map will not only change
the $Q$ system but also the reference system $R$ (see Eq.(\ref{wave1})). The
distortion is some kind of difference between the output and the input. If
the input state can not keep, the definition of the distortion will lost its
basis. Hence we can only define distortion on the basis of trace-preserving
Gaussian CP maps with a clearly physical meaning.

\textit{One mode Gaussian state: }The positivity of $\rho ^{RQ^{\prime }}$%
can be written as the uncertainty relation $\Gamma ^{\prime }-iJ_{12,n}\geq
0.$ Due to our selection of the purely off-diagonal $\beta $, we have $%
J_{12,n}=J_n\oplus (-J_n).$ For a one mode Gaussian state input, $\Gamma
^{\prime }$ is a $4\times 4$ matrix. The uncertainty relation requires $\det
(\Gamma ^{\prime }-iJ_{12})\geq 0$ which can be expressed as\cite{Simon}
\[
\det \Gamma _1^{\prime }\det \Gamma _2^{\prime }-\det \Gamma _1^{\prime
}-\det \Gamma _2^{\prime }+(1+\det \Gamma _{12}^{\prime })^2-Tr(J\Gamma
_1^{\prime }J\Gamma _{12}^{\prime }J\Gamma _2^{\prime }J\Gamma _{12}^{\prime
T})\geq 0.
\]
The sign before $\det \Gamma _{12}^{\prime }$ now is positive due to $%
J_{12}=J\oplus (-J).$ Denote $K=\det M,$ and $N=M^TN^{\prime }M,$ the
inequality will reduced to $K^2\det N^{\prime }-(1-K)^2\geq 0,$ that is
\begin{equation}
\det N-(1-K)^2\geq 0,  \label{wave}
\end{equation}
where we have used $\det (A+B)=\det A+\det B-Tr(JAJB),$ $M^TJM=K$ and $\det
\gamma _\psi =1.$ One the other hand, the positivity of the output state
reads $\Gamma _1^{\prime }-iJ\geq 0,$ thus $M^T\gamma M+N-iJ=M^T(\gamma
-iJ)M+N-iJ(1-K)\geq 0.$ For any input state we have $\gamma -iJ\geq 0,$ the
equality can be achieved by pure state, hence we have $N-iJ(1-K)\geq 0$
which will also lead to Ineq.(\ref{wave}).

The physical meaning of Gaussian trace-preserving map indicated by $M$ and $%
N $ is that $N$ is the additive noise and $M$ is a symplectic transformation
(rotation and squeezing) and a successively amplitude damping or
amplification. Let us consider the amplitude damping (as well as
amplification) of the channel, which is described by $\det (M)=K.$ The
amplitude of the signal is damped by a factor of $k=\sqrt{K}$, what will we
do to retrieve the input, clearly we will amplify it back. Or we will reduce
the input state with the same factor to compare with the output. In these
two cases, the distortion operators will be modified to $d(A,B)=\frac
12[(X_A/k-X_B)^2+(P_A/k+kP_B)^2]$ and $d(A,B)=\frac
12[(X_A-kX_B)^2+(P_A+kP_B)^2]$ respectively. In both these situations, if we
take all the steps as a whole channel, then we have $K=1.$ Hence in the
following we just need to consider the channel of symplectic transformation
and additive noise.

Let us consider the coherent information, which is determined by the
symplectic eigenvalues of $\Gamma ^{\prime }$ and $\Gamma _1^{\prime }.$ Now
$\det M=1,$ hence $(M\oplus \mathbf{I})$ are symplectic transformations,
which preserve the symplectic eigenvalues so that the coherent information. $%
\Gamma ^{\prime }$ can be written as $\Gamma ^{\prime }=(M^T\oplus \mathbf{I}%
)\Gamma ^{\prime \prime }(M\oplus \mathbf{I})$, with $\Gamma _1^{\prime
\prime }=\gamma +\left( M^T\right) ^{-1}NM^{-1},$ $\Gamma _{12}^{\prime
\prime }=\beta ,\Gamma _2^{\prime \prime }=\gamma $ correspondingly.

The problem is to search a $M$ such that the average distortion $\overline{d}
$ is minimized. We have
\begin{equation}
\overline{d}=\frac 14[TrN+TrM^T\gamma M+Tr\gamma -2TrM\beta \Lambda ].
\label{wave3}
\end{equation}
The minimization of $\overline{d}$ will involve algebra equation of power $%
4. $ Let us firstly consider the input of the thermal state whose CM is $%
\gamma _s\mathbf{I,}$ with $\gamma _s=2N_s+1$ is the symlectic eigenvalue of
the CM $\gamma $ and $N_s$ is the average photon number of the state. Denote
$\frac 14TrN=N_n$, we have
\[
M=\left[
\begin{array}{ll}
\sqrt{1+s^2-\delta ^2} & s+\kappa \\
s-\kappa & \sqrt{1+s^2-\delta ^2}
\end{array}
\right] ,
\]
with $s=\sqrt{\gamma _s^2-1}/(2\gamma _s)\equiv \sinh r_s$, and the
distortion operator
\[
\overline{d}=\frac 14(\frac 1{\gamma _s}+3\gamma _s)+N_n.
\]
The minimal $N_n$ is $0,$ hence the minimal $\overline{d}_{\min }=\frac
14(\frac 1{\gamma _s}+3\gamma _s),$ thus we define a canonical distortion $%
\overline{d}_c$ instead of $\overline{d}$, $\overline{d}_c=\overline{d}-%
\overline{d}_{\min },$%
\begin{equation}
\overline{d}_c=N_n.
\end{equation}
The coherent information now is determined by the symplectic eigenvalues of $%
\Gamma ^{\prime \prime }$ and $\Gamma _1^{\prime \prime }$ . The symplectic
eigenvalues are functions of the trace and determinant of the noise term $%
\left( M^T\right) ^{-1}NM^{-1}.$ Denote $\delta =\frac 14\det N,$ $\tau
=\frac 12Tr[\left( M^T\right) ^{-1}NM^{-1}],$ the coherent information of
the state with CM$\;\Gamma ^{\prime \prime }$ will be \cite{Chen}\cite
{Holevo}
\begin{equation}
I_c=g(d_0-\frac 12)-g(d_1-\frac 12)-g(d_2-\frac 12).
\end{equation}
Where $g(x)=(x+1)\log (x+1)-x\log (x)$ is the bosonic entropy function, and $%
d_0^2=x+(N_s+\frac 12)^2,$ $d_{1,2}^2=\frac 12[x+\frac 12\pm \sqrt{%
x^2-4N_s(N_s+1)\delta }],$ with $x=\delta +(N_s+\frac 12)\tau .$ The
entanglement information rate distortion $R^I$ now is the minimization of $%
I_c$ over all possible noise matrix $N$ with given trace $TrN=4N_n.$ After
the determinant and the trace of the noise matrix $N$ are given, we still
have the freedom in choosing the off-diagonal elements or the difference of
the diagonal elements. This freedom and the undetermined parameter $\kappa $
in the matrix $M$ are combined into a parameter $t$ ($-1\leq t\leq 1$) and
we can express $\tau $ as $\tau =2[N_n\cosh (2r_s)+t\sqrt{N_n^2-\delta }%
\sinh (2r_s)].$ Thus
\[
R^I=\min_{\delta ,t}I_c(x(\delta ,t),\delta ).
\]
The minimization will be achieved when $\delta =N_n^2,$ we prove this by
firstly preserving $x$ while increases $\delta .$ When $\sinh (2r_s)\geq
N_n/2,$ this is always possible by varying $t$ properly to compensate the
change of $x$ caused by the increase of $\delta $. We have $\frac{\partial
I_c(x,\delta )}{\partial \delta }=c_0[f(d_1-\frac 12)-f(d_2-\frac 12)].$
where $f(a)=\frac 1{2a+1}\log \frac{a+1}a$ is a monotonically decreasing
function and $c_0=\frac x{\sqrt{x^2-4N_s(N_s+1)\delta }}>0.$ Thus $\frac{%
\partial I_c(x,\delta )}{\partial \delta }\leq 0,$ $I_c(x,\delta )$
monotonically decreases with $\delta $ increases while preserving $x.$ The
minimum is achieved at $\delta =N_n^2$, that is, the noise matrix $N$ is
proportional to the unity matrix. The condition $\sinh (2r_s)\geq N_n/2$ may
contain most of the situations. For most of the input states (i.e. $%
N_s>0.012 $) when $N_n=2\sinh (2r_s),$ the values of the coherent
information will be $0$ at $\delta =N_n^2$. We need further to consider the
situation of weak signal input states (i.e. $N_s\leq 0.01$). There is the
case that $x(0,t)$ is too small compared with $x(N_n^2,t)$ for all $t.$ So
we need firstly increase $x$ from $x(0,t_0)$ to some intermediate state with
$x(\delta _1,t_1)=x(N_n^2,t)$ while in the $x$ increasing process the
coherent information is decreased. Let $x_{1,2}=\frac 12[x^2\pm
4N_s(N_s+1)\delta ]$, we increases $x$ and $\delta $ while keeping $x_2$
invariant, then $\frac{\partial I_c(x_1,x_2)}{\partial x_1}=\frac
1x[f(d_0-\frac 12)-\frac 12(f(d_2-\frac 12)+f(d_2-\frac 12))].$ The function
$f$ is not only a monotonically decreasing but also a downward convex
function. Hence in order to prove $\frac{\partial I_c(x_1,x_2)}{\partial x_1}%
\leq 0,$ we only need to prove $d_0\geq \frac 12(d_1+d_2)$ which is
confirmed by the fact that $d_0^2-\frac 12(d_1^2+d_2^2)$ $=$ $(N_s+\frac
12)^2+\frac 12x-\frac 14=N_s(N_s+1)+\frac 12x>0.$ This completes our proof.
The entanglement information rate distortion of thermal state input as a
function of the canonical distortion $N_n$ will be

\begin{equation}
R^I(N_n)=\max \{0,I_c(\delta =N_n^2,\tau =2N_n\cosh (2r_s))\}.
\end{equation}
where $\cosh (2r_s)=1+2N_s(N_s+1)/(2N_s+1)^2.$

For the general input $\gamma ,$ we may further transform the CM $\Gamma
^{\prime \prime }$ to $\Gamma ^{\prime \prime \prime }$ by symplectic
transformation $S\oplus S,$ $S$ diagonalizes $\gamma $ by $S^T\gamma
S=\gamma _s\mathbf{I}$, with $\gamma _s$ is the symplectic eigenvalue. Thus $%
\Gamma ^{\prime \prime \prime }$ will be a CM with its submatrices are $%
\Gamma _1^{\prime \prime \prime }=\gamma _s\mathbf{I}+S^T\left( M^T\right)
^{-1}NM^{-1}S,$ $\Gamma _{12}^{\prime \prime \prime }=\beta ,\Gamma
_2^{\prime \prime \prime }=\gamma _s\mathbf{I}$ correspondingly. $M$ is
determined by the minimization of $\overline{d}$. The above proving that
minimization of coherent information is achieved at $\delta =N_n^2$ remains
true for general input $\gamma $. The only difference is that we now have $%
\tau =[N_n\Omega +t\sqrt{N_n^2-\delta }\sqrt{\Omega ^2-4}],$ where $\Omega
=\sum_{i,j=1}^2M_{ij}^{\prime 2},$ $M^{\prime }=S^{-1}M.$ We have $\Omega
=\frac 1{\gamma _s}TrM^T\gamma M$ to be the function of trace and
determinant of the input CM $\gamma $. The entanglement information rate
distortion will be
\begin{equation}
R^I(N_n)=\max \{0,I_c(\delta =N_n^2,\tau =N_n\Omega )\}.
\end{equation}

A special case is the pure state input which contains squeezed states (for
coherent states and squeezed coherent states the distortion operator should
be modified). We have $\Omega =2$ which is the minimal of $\Omega .$ Thus $%
R^I(N_n)=\max \{0,g(N_s+N_n)-g(N_{sn1})-g(N_{sn2})\},$where $N_{sn1,2}=\frac
12(\sqrt{(N_n+1)^2+4N_nN_s}-1\pm N_n).$ Because $N_s=0$ so that $%
R^I(N_n)\equiv 0$ for pure states.

\begin{figure}[tbp]
\includegraphics[height=3in]{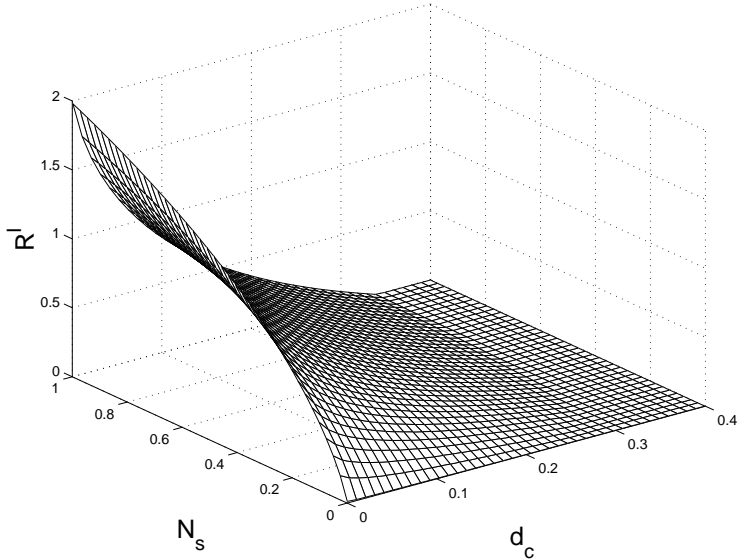}
\caption{Entanglement information rate distortion, the trace of input CM is $%
Tr\gamma=3, N_s=0$ for pure state, the maximal $N_s$ is for thermal state
input. }
\end{figure}

\textit{Conclusions and Discussions: }We proposed the distortion operator
which is quadratic of the canonical operators. The distortion operator has a
good classical correspondence of mean square error. The distortion is the
trace of the distortion operator on the joint state density operator. It is
an extension of the definition of classical distortion. For quantum Guassian
state source, we proved that the entanglement information rate distortion
which is a lower bound of the rate distortion is achieved by Gaussian map
under the constrain of zeroth, first and second moments. In the language of
distortion operator, distortion defined with fidelity ($1-F_e$) corresponds
to the distortion operator of $\mathbf{I-}\left| \Psi \right\rangle
\left\langle \Psi \right| ,$ where $\left| \Psi \right\rangle $ is the
purification of the source state. The quadratic canonical operator
distortion is more convenient than fidelity distortion for Gaussian state.

By the physical meaning of the distortion, we rule out the
non-trace-preserving Gaussian maps and convert the amplitude damping or
amplification channels to the standard maps which contain a symplectic
transformation and an additive noise in the language of correlation matrix.
For one-mode Gaussian state input, we proved that the entanglement
information rate distortion is achieved when the additive noise matrix is
proportional to unity matrix. The canonical distortion is simply the average
photon number of the noise. The rate distortion for pure state input is zero.

One of the most important conclusion we can draw is that the rate distortion
function is accessible for noiseless case. For any one mode Gaussian input
states, the entanglement information rate distortion functions at the point
of zero distortion are $R^I(0)=g(N_s),$ which is the entropy of the source $%
S(\rho ).$ From Schumacher's quantum noiseless coding theorem\cite
{Schumacher} we know that $R(0)=S(\rho ).$ Thus we have the conclusion that $%
R(0)=R^I(0).$

\textit{Acknowledgement: }Funding by the National Natural Science Foundation
of China (under Grant No. 10575092), Zhejiang Province Natural Science
Foundation (Fund for Talented Professionals, under Grant No. RC104265) and
AQSIQ of China (under Grant No. 2004QK38) are gratefully acknowledged.

\end{document}